\documentclass{article}

\usepackage{arxiv}

\usepackage[utf8]{inputenc} 
\usepackage[T1]{fontenc}    
\usepackage{hyperref}       
\usepackage{url}            
\usepackage{booktabs}       
\usepackage{amsfonts}       
\usepackage{nicefrac}       
\usepackage{microtype}      
\usepackage{lipsum}		
\usepackage{graphicx}
\usepackage[numbers,sort&compress]{natbib}
\usepackage{doi}
\usepackage{siunitx}

\title{Scaling laws to predict humidity-induced swelling and stiffness in hydrogels}

\author{ 
	Yiwei Gao \\
	Department of Mechanical Engineering\\
	University of Nevada, Las Vegas\\
	Las Vegas, NV 89154 \\
	\And
	Nicholas K.K. Chai \\
	Department of Mechanical Engineering\\
	University of Nevada, Las Vegas\\
	Las Vegas, NV 89154 \\
	\And
	Negin Garakani \\
	Department of Mechanical Engineering\\
	University of Nevada, Las Vegas\\
	Las Vegas, NV 89154 \\
	\And
	\href{https://orcid.org/0000-0003-2400-1561}{\includegraphics[scale=0.06]{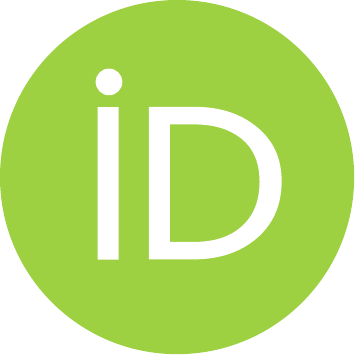}\hspace{1mm}Sujit S. Datta}\thanks{\href{https://dattalab.princeton.edu/}{https://dattalab.princeton.edu/}} \\
	Department of Chemical and Biological Engineering\\
	Princeton University\\
	Princeton, NJ 08544 \\
	\texttt{ssdatta@princeton.edu} \\
	\And
	\href{https://orcid.org/0000-0001-5857-0055}{\includegraphics[scale=0.06]{orcid.pdf}\hspace{1mm}H. Jeremy Cho}\thanks{\href{https://dakine.sites.unlv.edu/}{https://dakine.sites.unlv.edu/}} \\
	Department of Mechanical Engineering\\
	University of Nevada, Las Vegas\\
	Las Vegas, NV 89154 \\
	\texttt{jeremy.cho@unlv.edu} \\
}

\renewcommand{\shorttitle}{Scaling laws to predict humidity-induced swelling and stiffness in hydrogels}

\hypersetup{
pdftitle={Scaling laws to predict humidity-induced swelling and stiffness in hydrogels},
pdfauthor={Yiwei Gao, Nicholas K.K. Chai, Negin Garakani, Sujit S. Datta, H. Jeremy Cho},
}

\begin{document}
\maketitle

\begin{abstract}
	From pasta to biological tissues to contact lenses, gel and gel-like materials inherently soften as they swell with water. In dry, low-relative-humidity environments, these materials stiffen as they de-swell with water. Here, we use semi-dilute polymer theory to develop a simple power-law relationship between hydrogel elastic modulus and swelling. From this relationship, we predict hydrogel stiffness or swelling at arbitrary relative humidities. Our close predictions of properties of hydrogels across three different polymer mesh families at varying crosslinking densities and relative humidities demonstrate the validity and generality of our understanding. This predictive capability enables more rapid material discovery and selection for hydrogel applications in varying humidity environments.
\end{abstract}


\section{Introduction}
Hydrogels, which are polymer networks that absorb water, have attracted increased attention in recent decades due to their distinct water-holding behavior. They have been shown to be beneficial for a variety of applications from horticulture \cite{kim2010polyacrylamide,kalhapure2016hydrogels,johnson1984effects,louf2021under} to soft robotics or tissue engineering \cite{zhou2019highly,subramani2020influence,liu2020programmable,lin2016stretchable,jiang2011pva}. Hydrogels used for soft robotics and actuation rely on the inherent changes in mechanical stiffness that result from changes in the amount of water in the hydrogel. This relationship is observed in everyday foods such as rice or pasta—which can be described as starch-based hydrogels—wherein softness increases with water content. In the past three decades, there has been a large body of research focusing on either the swelling behavior \cite{kim2010polyacrylamide,manjula2013preparation,bajpai2001swelling,isik2004preparation,ccaykara2003network,icsik2004swelling,stanojevic2006investigation,saraydin2004influence} or the mechanical stiffness \cite{zhou2019highly,subramani2020influence,jiang2011pva,zahouani2009characterization,shetye2015hydrogels,yang2012mechanical,huang2007novel,normand2000new,mongia1996mucoadhesive,cohen1992characterization,sun2012highly,li2014hybrid,meyvis2000influence,muniz2001compressive,wyss2010capillary,ming2020switching,zrinyi1987elastic,ahearne2005characterizing,sun2012highly}. One notable example is the study by Li et al., which used Flory-Huggins theory to develop an equation of state that could be used to relate swelling to osmotic pressure \cite{li2012experimental}. In particular, they found that osmotic pressure was independent of crosslinking density, indicating that gels composed of the same base monomer can be treated similarly. Their work, and the collective work of others, demonstrates a strong fundamental understanding of hydrogel swelling and stiffness behavior in conditions close to a fully swollen state; however, we have a less-developed understanding of how swelling and stiffness depend on humidity. Hydrogel studies on stiffness often limit analysis to the fully wet state \cite{cohen1992characterization,sun2012highly,li2014hybrid,cong2014highly,muniz2001polyacrylamide,li2014stiff,denisin2016tuning}. Water swelling in hydrogels is also controlled by the relative humidity in the ambient environment. This humidity-induced swelling is particularly important for food preservation and preparation\cite{basu2006models}. Recent works in atmospheric water harvesting have relied on hygroscopic sorbents—including gels \cite{kallenberger2018water,matsumoto2018thermo,zhao2019super}—that absorb water at different humidities\cite{zhou2020atmospheric}. The moisture sorption isotherm quantifies how much water these materials can absorb—or swell in the case of gels. As humidity increases, hydrogels should swell and soften; however, the exact dependence on humidity remains an open question. Here, we present simple scaling laws based on semi-dilute polymer theory that (1) describe the dependence of stiffness on swelling and (2) dependence of stiffness and swelling on relative humidity. Using these scaling laws, we present a method to predict moisture sorption isotherms from limited stiffness and sorption data from a similar reference hydrogel.
\section{Results}
Starting from semi-dilute polymer theory, we develop a scaling law relationship between mechanical bulk modulus and swelling fraction. We then relate this dependency to changes in osmotic pressure and relative humidity.
\subsection{Dependence of stiffness on swelling}
To develop a direct relationship between elastic modulus and swelling, we use de Gennes’ semi-dilute description of polymer solutions \cite{de1979scaling}. The utility of de Gennes’ semi-dilute description is such that many properties of polymer solutions can be quantified using simple power-law relationships. In accordance with numerous studies \cite{sakai2012effect,bhattacharyya2020hydrogel,schulze2017polymer}, we assume the hydrogel can be thought of as a semi-dilute solution where the monomer concentration, $c$, is slightly higher than the overlap concentration, $c^*$. At this semi-dilute state, polymer “blobs” are entangled with each other, creating an expansive polymer mesh. This polymer mesh is characterized by an average spacing between polymer chains termed the correlation length, $\xi$. de Gennes determined that the correlation length is 
\begin{equation}
    \xi = a^{7/4}v^{-1/4}\phi_{\text{poly}}^{-3/4} \label{eq1}
\end{equation}
where $a$ is the monomer size, $v$ is the excluded volume of the monomer. $\phi_{\text{poly}}$ is the volume fraction of polymer defined as $\phi_{\text{poly}} \equiv V_{\text{poly}}/V$ where $V$ is the volume of the material including solvent and $V_\text{poly}$ is the volume of polymer exlucing solvent. The affinity between monomer and solvent is captured by the excluded volume term as it is related to the Flory-Huggins interaction parameter, $\chi$, such that $v = a^3(1-2\chi)$.
\begin{figure}[htb]
    \centering
    \includegraphics[width=\linewidth]{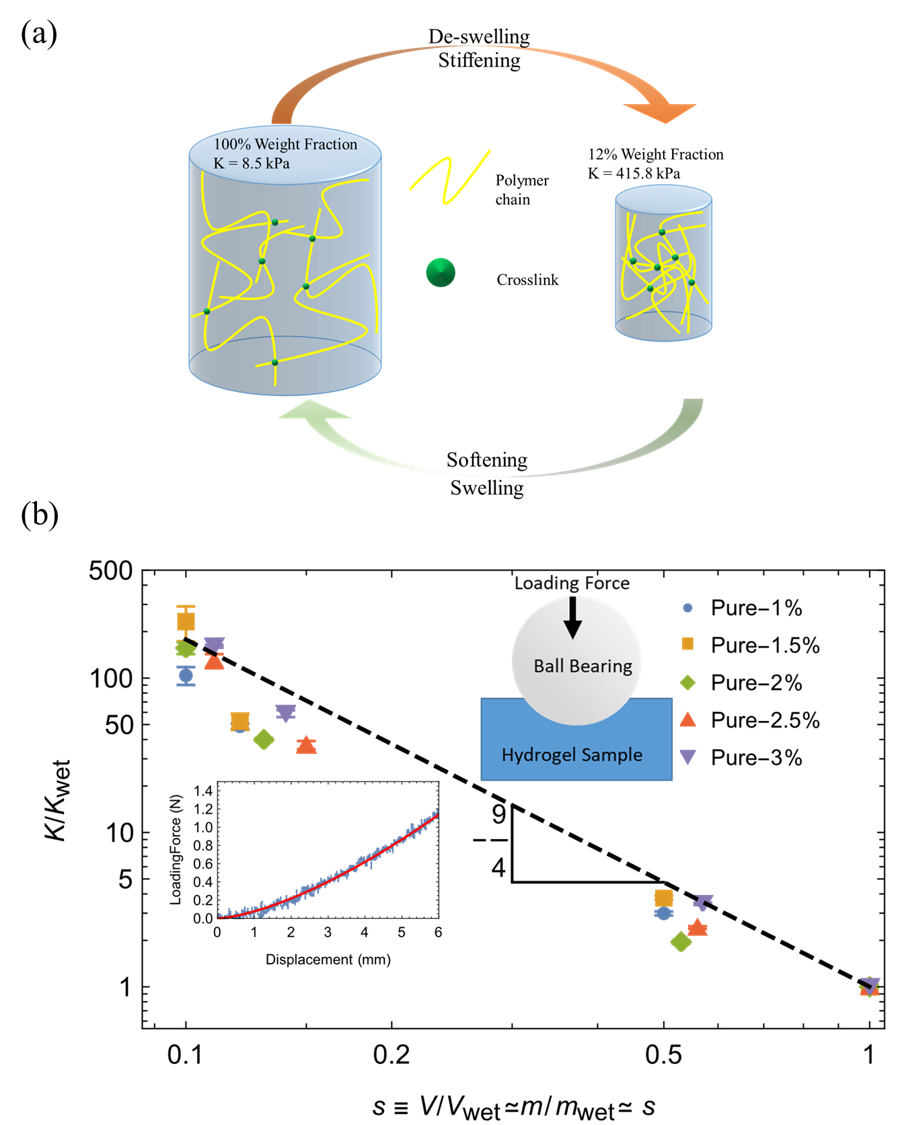}
    \caption{There is an inherent relationship between swelling and stiffness. (a) As a hydrogel de-swells, the crosslinked polymer mesh densifies and stiffens. In (a), we show swelling and stiffness data for \SI{1}{\percent} crosslinked pure PAAm. (b) Across a range of crosslinkings from \SIrange{1}{3}{\percent}, we found that stiffness depends on the $-9/4$ power with swelling fraction. (b,inset) The stiffness, as quantified by the bulk modulus, is determined using contact mechanics and force-displacement data from indentation tests. When moduli are normalized by their reference wet-state value, $K_\text{wet}$, all stiffness and swelling fraction data collapse onto a single curve according to Eq.~\ref{Eq.5}. Error bars represent uncertainties in repeatability and curve fitting (see Materials and Methods).}
    \label{fig.1}
\end{figure}
The osmotic pressure describes the state of swelling in a polymer solution and is directly related to the volume fraction of the polymer, $\phi_{\text{poly}}$, where higher $\phi_{\text{poly}}$ results in higher osmotic pressures. As such, hydrogels that are more swollen (low $\phi_{\text{poly}}$) have lower osmotic pressures. de Gennes showed that the osmotic pressure, $\Pi$, is related to the correlation length such that $\Pi = CkT/\xi^3$, where $C$ is a dimensionless constant on the order of unity and $kT$ is the product of the Boltzmann constant and absolute temperature. Applying Eq.~1 to this relation, the osmotic pressure can be expressed as a scaling law with $\phi_{\text{poly}}$ such that 
\begin{equation}
    \Pi = CkTa^{-21/4}v^{3/4}\phi_{\text{poly}}^{9/4}. \label{Eq.2}
\end{equation} 

If we take a gel and compress or expand it, we are changing its volume, $V$, without changing the number of monomers. Assuming that a hydrogel can be considered a poroelastic system \cite{louf2021poroelastic}, the elastic modulus of interest is the drained bulk modulus since it allows solvent to drain in and out freely while the polymer structure is compressed or expanded. We denote this modulus as $K$, which is defined as $K \equiv -V{\partial{P_{\text{ext}}}}/{\partial{V}}$ where $P_{\text{ext}}$ is the external pressure applied on the body. At chemical and mechanical equilibrium, the external pressure can be related to the osmotic pressure such that $P_{\text{ext}} = P_{\text{0}}+\Pi$ where $P_{\text{0}}$ is the ambient pressure, assumed to be a constant. Thus, 
\begin{equation}
    K = -V\frac{\partial\left(P_{\text{0}}+\Pi\right)}{\partial{V}} = -V\frac{\partial{\Pi}}{\partial{V}} \label{Eq.3}
\end{equation} 
indicating that the elastic modulus of a polymer solution is related to changes in osmotic pressure. Applying de Gennes’ power-law expression of osmotic pressure (Eq.~2) in this definition of modulus,
\begin{equation}
    K \sim kTa^{-21/4}v^{3/4}\phi_{\text{poly}}^{9/4}. \label{Eq.4}
\end{equation} 
This result shows that the stiffness of a hydrogel scales with the volume fraction of polymer to the ${9/4}$ power. Therefore, from the wet (fully swollen) to drier states, the volume fraction of polymer, $\phi_{\text{poly}}$, increases, and the hydrogel stiffens as a result as long as the semi-dilute description holds. Eq.~4 allows for determination of $K$ from measurable quantities $T$, $a$, $v$ and $\phi_{\text{poly}}$. However, determining these quantities involves many separate, time-consuming experimental procedures. It is considerably more convenient to deal with a scaling law that depends on a reference state that can be readily characterized with fewer experiments. Therefore, we develop a reduced scaling law for stiffness using a reference wet-state modulus $K_{\text{wet}}$. At this reference wet state, the hydrogel is swollen and in equilibrium with pure water or, equivalently, \SI{100}{\percent} humidity. We can define a swelling fraction as $s \equiv V/V_{\text{wet}}$ where $V_{\text{wet}}$ is the volume of the gel at the wet state and $V$ is the volume at an arbitrary state of swelling, i.e., equilibrated at an arbitrary relative humidity. Note that the swelling fraction has a maximum value of unity since at the maximum swelling, $V=V_\text{wet}$. Conversely, at the driest possible state, $V = V_\text{poly}$ and, therefore, $s=V_\text{poly}/V_\text{wet}<1$. Thus, the range of swelling fraction is $V_\text{poly}/V_\text{wet} \leq s \leq 1$. While there are related quantities to the swelling fraction---e.g., the swelling ratio $J \equiv V/V_\text{poly}$ \cite{li2012experimental} and the degree of swelling $\Phi \equiv V_\text{wet}/V$ \cite{Cho2019prl,Cho2019sm}---the swelling fraction more intuitively describes the fractional content of water based off of a reference wet state at \SI{100}{\percent} humidity. Taking the definition of polymer volume fraction, $\phi_{\text{poly}} \equiv V_{\text{poly}}/V$, and eliminating $V$ using the definition of swelling fraction, we obtain $\phi_{\text{poly}} = \frac{V_{\text{poly}}}{V_{\text{wet}}s}$. Substituting this expression for $\phi_{\text{poly}}$ into Eq.~4, we find that $K \sim kTa^{-21/4}v^{3/4}(\frac{V_{\text{wet}}}{V_{\text{poly}}}s)^{-9/4}$. Finally, using the ratio of moduli at the arbitrary and wet swelling states, $K/K_{\text{wet}}$, the constant factor of $kTa^{-21/4}v^{3/4}$ cancels out and a reduced modulus can be related to the swelling fraction as
\begin{equation}
    \frac{K}{K_{\text{wet}}} = \frac{\left(\frac{V_{\text{wet}}}{V}\right)^{9/4}}{\left(\frac{V_{\text{wet}}}{V_{\text{wet}}}\right)^{9/4}} = 
    \left(\frac{V_{\text{wet}}}{V}\right)^{9/4} = s^{-9/4}.\label{Eq.5}
\end{equation} 
Thus, as a gel de-swells (decreasing $s$), the modulus sharply increases due to the collapse of the polymer network as illustrated in Fig.~1a.
\begin{figure*}
    \centering
    \includegraphics[width=\textwidth]{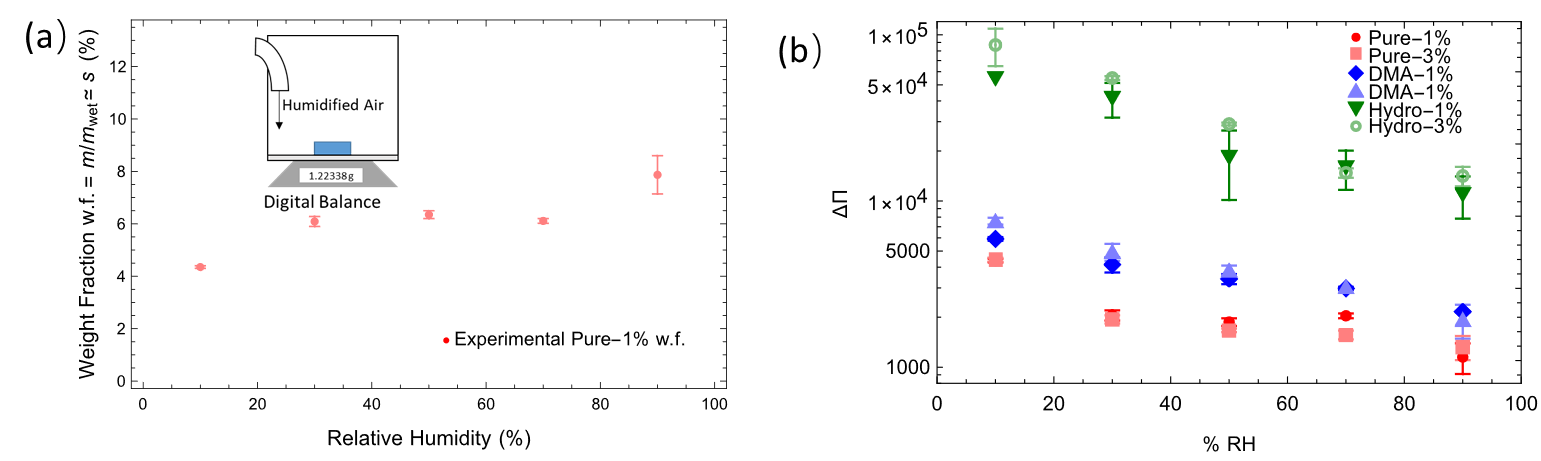}
    \caption{Lowering relative humidity, (a) de-swells hydrogels as quantified by the moisture sorption isotherm and (b) increases osmotic pressure. Hydrogel samples were equilibrated at different relative humidities ranging from \SIrange{10}{90}{\percent}. In (b), we use swelling fraction and wet-state stiffness to calculate $\Delta\Pi = \frac{4}{9}K_{\text{wet}}(s^{-9/4}-1)$ (Eq.~\ref{Eq.6}). We show only samples with crosslinking ratios of \SI{1}{\percent} and \SI{3}{\percent}, representing the entire range of crosslinking, for clarity. Error bars in weight fraction represent measurement and repeatability uncertainties. Error bars in $\Delta\Pi$ represent propagation errors in Eq.~\ref{Eq.6} originating from uncertainties in weight fraction and $K_\text{wet}$.}
    \label{fig.2}
\end{figure*}

To verify this relationship between swelling and stiffness (Eq.~5), we perform mechanical indentation tests to measure the elastic modulus at various states of swelling (see Materials and Methods)—a verified mechanical characterization technique for soft gels\cite{schulze2017polymer}. By indenting a sample with a spherical indenter and measuring its force-displacement response, we apply Hertzian contact mechanics \cite{Johnson1985} to determine the elastic modulus (see Materials and Methods). We set the timescales of our measurements to ensure that we measure the drained bulk modulus. Our results do not depend on displacement speed (in our range of testing), indicating that the material behaves quasi-statically and quasi-elastically, away from any dynamic drainage or viscoelastic effects.

 We use N,N'-methylene(bis)acrylamide (MBA)-crosslinked PAAm hydrogels at different crosslinking ratios (\SI{1}{\percent}, \SI{1.5}{\percent}, \SI{2}{\percent}, \SI{2.5}{\percent}, and \SI{3}{\percent} (mol MBA)/(mol PAAm) \%). We achieve different states of swelling by hydrating and dehydrating samples at different relative humidities; the wet state is achieved by equilibrating samples in water. In accordance with other work \cite{icsik2004swelling,jayaramudu2019swelling,skelton2013biomimetic}, we assume similar densities between polymer and water and a relatively larger amount of water compared to polymer such that $s$ can be determined from a ratio of weights $s \simeq m/m_{\text{wet}}$. We find that across approximately one decade of swelling fraction, the measured elastic modulus closely followed the $-9/4$ scaling law as predicted by Eq.~5 for all crosslinking densities tested, as shown in Fig.~1b. This close agreement is consistent with previous studies \cite{zrinyi1987elastic,obukhov1994network,bell1995biomedical} and verifies our two assumptions that (1) the hydrogel network can be represented as a semi-dilute polymer solution and (2) the hydrogel network is a mechanical poroelastic system where the elastic (drained bulk) modulus is equivalent to the osmotic modulus regardless of crosslinking density.

\subsection{Dependence of stiffness and swelling on relative humidity}
 The relative humidity of the ambient environment sets the chemical potential of water vapor. A gel that is equilibrated with the ambient will have an internal chemical potential equivalent to the ambient set by the relative humidity: $\mu
 = \mu_\text{wet} + kT\ln\left(\text{RH}\right)$ where the wet state corresponds to \SI{100}{\percent} humidity and $\text{RH}$ is the fractional relative humidity so $ kT\ln\left(\text{RH}\right) \leq 0$. We use the osmotic pressure as a mediating variable to understand the dependence of stiffness and swelling on relative humidity. Since osmotic pressure is a volumetric form of chemical potential where $\Pi \approx \frac{\mu_0 - \mu}{v_\text{w}}$ ($\mu_0$ being the chemical potential of pure solvent and $v_\text{w}$ being an effective molecular volume that depends on polymer type), it also varies with relative humidity:
\begin{equation}
    \Pi \approx \frac{\mu_0 - \mu_\text{wet} - kT\ln\left(\text{RH}\right)}{v_\text{w}} \label{eqn:pirh}
\end{equation}
such that osmotic pressure increases with decreasing relative humidity.
Conversely, as relative humidity increases, the osmotic pressure decreases, and the gel swells. To understand this from the quantity of swelling fraction, according to Eq.~2 and the relationship between $\phi_\text{poly}$ and $s$, the scaling relationship between osmotic pressure and swelling fraction is $\Pi \propto s^{-9/4}$. The relationship between $s$ and $\text{RH}$ is the moisture sorption isotherm as it quantifies the amount of water absorbed as a function of humidity. As shown in Fig.~2a, the moisture sorption isotherm can be experimentally determined for any gel using mass measurements of samples equilibrated to arbitrary relative humidities.

When $\text{RH}<1$ and gels de-swell from their wet state, $s$ decreases, and the osmotic pressure experiences a change of $\Delta\Pi \equiv \Pi - \Pi_{\text{wet}}$. Using Eq.~\ref{eqn:pirh}, we can quantify changes in osmotic pressure with relative humidity as
\begin{equation}
\Delta\Pi \approx -kT\ln\left(\text{RH}\right)/v_\text{w}\text{.} \label{deltapiapprox}
\end{equation}
We can also quantify $\Delta\Pi$ in terms of $s$ and the mechanical stiffness at the wet state, $K_\text{wet}$. To do so, we apply the definition of modulus, $K$ (Eq.~3), into our reduced scaling law (Eq.~5) and obtain $\frac{-V\frac{\partial{\Pi}}{\partial{V}}}{K_{\text{wet}}} = (\frac{V}{V_{\text{wet}}})^{-\frac{9}{4}}$. Integrating both sides from the wet state to an arbitrary swelling state $\frac{-1}{K_{\text{wet}}}\int_{\Pi_{\text{wet}}}^{\Pi} \,\partial{\Pi'}\ = \int_{V_{\text{wet}}}^{V} \frac{1}{V'}(\frac{V'}{V_{\text{wet}}})^{-\frac{9}{4}} \,\partial{V'}\,$, we find that the change in osmotic pressure from the wet state, $\Delta\Pi \equiv \Pi - \Pi_{\text{wet}}$ is
\begin{equation}
    \Delta\Pi = \frac{4}{9}K_{\text{wet}}(s^{-9/4}-1). \label{Eq.6}
\end{equation} 
From this relationship, we observe that $\Delta\Pi$ can be calculated from a mechanical measurement at the wet state and the swelling fraction at any abitrary swelling state at particular $\text{RH}$. Plotting this calculated $\Delta\Pi$ with $\text{RH}$ we observe that $\Delta\Pi$ increases with decreasing $\text{RH}$ as shown in Fig.~2b. Alternatively, from Eq.~\ref{deltapiapprox}, $\Delta\Pi$ only depends on $\text{RH}$ and the effective molecular volume, $v_\text{w}$. Thus, we expect that hydrogels of similar $v_\text{w}$ should experience similar changes in osmotic pressure.

\begin{figure}
    \centering
    \includegraphics[width=\linewidth]{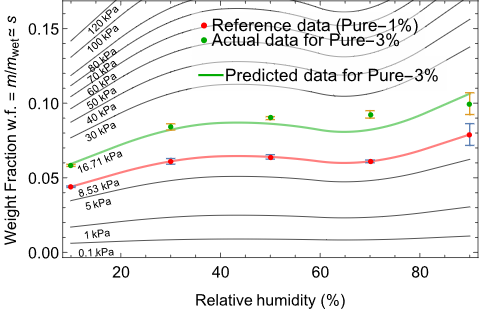}
    \caption{Using a known moisture sorption isotherm for a reference hydrogel (pure PAAm \SI{1}{\percent}, red) with known wet-state stiffness, we can predict moisture sorption isotherms (black) of similar hydrogels with any arbitrary wet-state stiffness, $K_\text{wet}$, using Eq.~\ref{Eq.6} and ~\ref{Eq.7}. There is close agreement between our prediction and an experimentally obtained moisture sorption isotherm for a similar hydrogel with $K_\text{wet}=\SI{16.7}{\kilo\pascal}$ (pure PAAm \SI{3}{\percent}, green). Error bars in weight fraction represent measurement and repeatability uncertainties.}
    \label{fig.3}
\end{figure}
\subsection{Equivalence of osmotic pressure differences across similar hydrogels}
Motivated by Li et al.’s determination that hydrogels of the same polymer composition have the same osmotic pressures, independent of crosslinking density \cite{li2012experimental}, we expect that similar hydrogels that differ only by crosslinking density have similar $v_\text{w}$. Therefore, from Eq.~\ref{deltapiapprox}, gels differing only by crosslinking experience the same changes in osmotic pressure, $\Delta\Pi$, when exposed to the same relative humidities. Indeed, the results shown in Fig.~2b confirm that  the $\Delta\Pi$ values are independent of crosslinking density within uncertainty (red data points for pure PAAm hydrogels). To further confirm this, we tested two other polymer meshes, each at different crosslinking densities: PAAm with the addition of N,N-Dimethylacrylamide (DMA) \cite{aalaie2008gelation}; and PAAm post-treated with hydrolysis using sodium hydroxide\cite{kim2010polyacrylamide}. Like pure PAAm, both PAAm + DMA (blue) and hydrolyzed PAAm (green) samples achieved the same $\Delta\Pi$ at the same humidities, independent of crosslinking density. FTIR spectra (Fig.~S4) for each hydrogel indicate three different hydrogel polymer mesh families and independence with crosslinking density, corroborating our findings.

\subsection{Predicting moisture sorption isotherms of similar hydrogels}
From the equivalence in $\Delta\Pi$ across different crosslinking densities, we can confirm that
\begin{equation}
    \Delta\Pi(\text{RH})_{\text{A}} = \Delta\Pi(\text{RH})_{\text{B}} \label{Eq.7}
\end{equation} 
for any two samples (sample A and sample B) of the same polymer mesh and relative humiditiy. Setting two $\Delta\Pi$ expressions equal to each other using Eq.~\ref{Eq.6} for two similar hydrogels with different crosslinkings enables one to predict unknown properties of one of the hydrogels. For instance, we can predict the moisture sorption isotherm—water weight fraction, $\text{w.f.} \simeq s$, versus $\text{RH}$ for sample B as long as we know the moisture sorption isotherm of a reference hydrogel A, the reference wet-state stiffness, $K_{\text{wet, A}}$, and the wet-state stiffness of hydrogel B, $K_{\text{wet, B}}$. Furthermore, using the scaling law for modulus and swelling fraction (Eq.~\ref{Eq.5}), we can calculate stiffness of hydrogel B at any relative humidity. Thus, one only needs to study a specific reference hydrogel in detail to understand the humidity-dependent swelling and stiffness of an entire family of hydrogels.

\begin{figure}[htb]
    \centering
    \includegraphics[width=\linewidth]{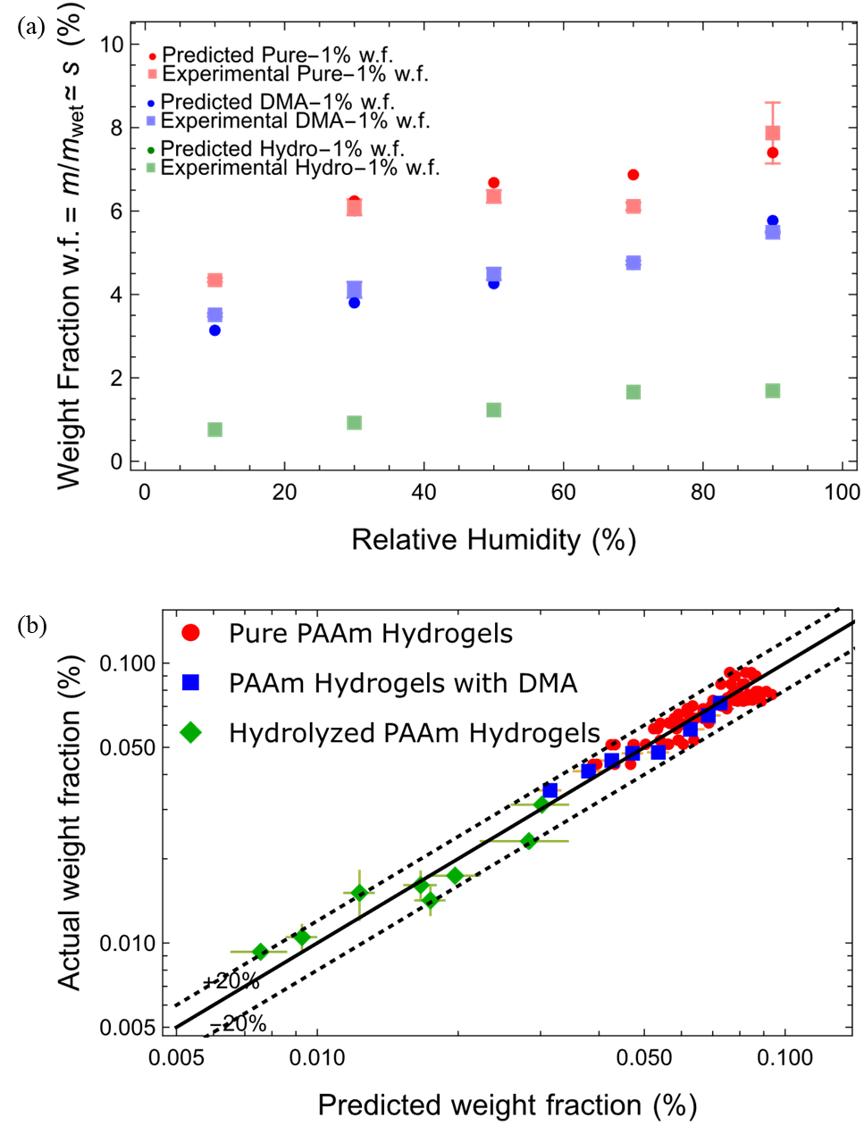}
    \caption{(a) The close agreement between predictions of moisture sorption isotherms across three different hydrogel families (pure PAAm, PAAm + DMA, hydrolyzed PAAm) verifies our scaling laws and prediction scheme. We show only data for \SI{1}{\percent} crosslinker ratio samples for clarity. We further verify the accuracy of our approach with (b) over 120 separate verifications of hydrogel samples across these three families at all crosslinker ratios (\SI{1}{\percent}, \SI{1.5}{\percent}, \SI{2}{\percent}, \SI{2.5}{\percent}, and \SI{3}{\percent}) and humidities (\SI{10}{\percent}, \SI{30}{\percent}, \SI{50}{\percent}, \SI{70}{\percent}, and \SI{90}{\percent}) where we found close agreement between predictions and measurements of weight fraction within $\pm\SI{20}{\percent}$. Error bars in actual weight fraction represent measurement and repeatability uncertainties. Error bars in predicted weight farctions represent propagation errors in Eq.~\ref{procedureeqn} originating from uncertainties in weight fraction and $K_\text{wet}$.}
    \label{fig.4}
\end{figure}

Using a pure PAAm hydrogel at \SI{1}{\percent} crosslinking as a reference (sample A) where we know $s_A$ at any relative humidity and its $K_\text{wet,A}$, we demonstrate the procedure to predict the sorption isotherm of a related hydrogel: pure PAAm hydrogel at \SI{3}{\percent} crosslinking (sample B). We start by setting the two $\Delta\Pi$ terms equal to each other (Eq.~\ref{Eq.7}) and expressing them in terms of their respective $K_\text{wet}$ and $s$ (Eq.~\ref{Eq.6}). Given a known reference sample swelling fraction, $s_A$, at a particular humidity, we need to determine the unknown $s_B$ at the same humidity. By performing a single mechanical test to determine $K_\text{wet,B} = \SI{16.7}{\kilo\pascal}$ and applying the known $K_\text{wet,A} = \SI{8.5}{\kilo\pascal}$, we can use the equivalence of $\Delta\Pi$ to determine the single unknown $s_B$:
\begin{equation}
    \frac{4}{9}\underbrace{K_{\text{wet,A}}}_{\text{known}}\left({\underbrace{s_\text{A}}_{\text{known}}}^{-9/4}-1\right) = \frac{4}{9}\underbrace{K_{\text{wet,B}}}_{\text{known}}\left({\underbrace{s_\text{B}}_{\text{unknown}}}^{-9/4}-1\right) \label{procedureeqn}
\end{equation}

This procedure works to determine the swelling fraction $s_B$ at a particular $\text{RH}$. Repeating this procedure for any $\text{RH}$, we can determine the entire moisture sorption isotherm for sample B as shown by the green curve in Fig.~\ref{fig.3}. For this particular case, the predicted sorption isotherm for B is in close agreement with an experimentally determined sorption isotherm (Fig.~\ref{fig.3}, green dots). Furthermore, the moisture sorption isotherm for any hydrogel similar to A with arbitrary wet-state stiffness, $K_\text{wet}$, can be determined (black curves in Fig.~\ref{fig.3}).

The equivalence of $\Delta\Pi$ across similar hydrogels can also be used to determine stiffness at any arbitrary relative humidity. If $s_B$ is known at a particular humidity below \SI{100}{\percent} and $K_\text{wet,B}$ is unknown, then we can use Eq.~\ref{procedureeqn} to determine $K_\text{wet,B}$. Having determined $K_\text{wet,B}$, Eq.~\ref{procedureeqn} can subsequently be used to determine $s_B$ at any relative humidity with A as a reference. Having obtained $s_B$ as a function of humidity, we can then apply the stiffness–swelling law, Eq.~\ref{Eq.5}, to determine $K_\text{B}$ at any arbitrary humidity.

To verify our prediction approach, we perform 120 independent weight fraction predictions for nine different hydrogels from three families, each at five different humidities (full results in Fig.~S1). For each prediction at a particular humidity, we use different samples as references. For example, to predict the weight fraction of pure–\SI{2.5}{\percent} at \SI{50}{\percent} $\text{RH}$, we use pure PAAm at \SI{1}{\percent},  \SI{1.5}{\percent}, \SI{2}{\percent}, and \SI{3}{\percent}, all at \SI{50}{\percent} $\text{RH}$, as references, representing four independent predictions for pure–\SI{2.5}{\percent} at \SI{50}{\percent} $\text{RH}$. These multiple predictions are possible by performing moisture sorption experiments across a range of humidities for every sample. This dataset also serves as a means to verify our predictions. Comparing our predictions and actual measurements of weight fraction, we find close agreement as shown in Fig.~4a. In Fig.~4b, we compare our predictions to experimentally measured weight fraction values and find that they are accurate to within $\pm 20\%$. In fact, more than $70\%$ of the samples were accurate to within $\pm10\%$. The remarkably close agreement across a wide range of hydrogels provides strong validation of (1) the assumption that osmotic pressure changes are equivalent for similar hydrogels at the same humidities and (2) the scaling law relationship between stiffness and swelling derived from semi-dilute polymer theory. 

\section{Conclusions}
Our work shows that semi-dilute polymer theory can be applied to develop a simple power-law relationship between swelling and stiffness. We also elucidate how swelling and stiffness depend on relative humidity using the concept of osmotic pressure. We find that changes in osmotic pressure due to humidity changes are equivalent across similar hydrogels, indpendent of crosslinking. Combining the stiffness–swelling power law with the the principle equivalent changes in osmotic pressure, we demonstrate procedures to predict swelling or stiffness at any relative humidity for any arbitrary hydrogel of similar polymer network. With our prediction procedures, one only needs to study a specific reference hydrogel in detail to understand the humidity-dependent swelling and stiffness of any related hydrogels within the same family. We define hydrogel similarity using the effective molecular volume, $v_\text{w}$, which relates osmotic pressure to chemical potential. Further investigation of the molecular interactions that affect $v_\text{w}$ could add to the predictive power of our approach and inform how we could classify hydrogel families. The simplicity of our resulting equations provides substantial utility for gel synthesis design. We anticipate that our work will guide hydrogel applications such as agriculture and soft robotics that depend on the inherent relationship between swelling and stiffness, operating at different humidities.

\section{Materials and methods}
\subsection{Preparation of hydrogels}
All hydrogels were prepared from aqueous stock solutions of the following chemicals: N,N'-methylene(bis)acrylamide (MBA), N,N-Dimethylacrylamide (DMA), ammonium persulfate (APS), and tetramethylethylenediamine (TEMED) at concentrations of \SI{0.127}{\gram}/\SI{10}{\milli\liter}, \SI{2.6}{\milli\liter}/\SI{10}{\milli\liter}, \SI{0.08}{\gram}/\SI{10}{\milli\liter} and \SI{0.25}{\milli\liter}/\SI{10}{\milli\liter}, respectively. The base acrylamide (AAm) monomer was used in its pure powder form. By mixing different amounts of these chemicals, polymers were spontaneously synthesized. During this process, APS served as an initiator, TEMED as an accelerator, and MBA as a crosslinker. In all hydrogels, we started with \SI{0.25}{\gram} of AAm monomer, \SI{0.5}{\milli\liter} of TEMED solution and \SI{0.5}{\milli\liter} of APS solution. Then, we mixed varying amounts of MBA solution in order to achieve the target crosslinker ratios (moles of MBA over moles of AAm) ranging from \SIrange{1}{3}{\percent}. To ensure that polymerization occurred in a consistently dilute environment for all hydrogels, we added DI water to ensure that the mole fraction of water over other chemicals was 1000. For hydrogels with DMA, \SI{20}{\percent} (DMA/AAm mol/mol) was added. Then, the solution was vortex mixed for approximately one minute and subsequently rested at room temperature (\SI{24}{\celsius}) for 24 hours. For samples that were hydrolyzed, we immersed the samples in \SI{1}{\mole\per\liter} sodium hydroxide for 30 minutes before. Finally, the samples were flushed in DI water for one week to remove unreacted chemicals and equilibrate them to the wet state.

The samples are named by their method of treatment and crosslinker ratio. For example, the pure PAAm hydrogel with \SI{1}{\percent} crosslinker ratio is named Pure-\SI{1}{\percent}; the DMA-modified hydrogel with \SI{2}{\percent} crosslinker ratio is named DMA-\SI{2}{\percent}; the hydrolyzed hydrogel with \SI{3}{\percent} crosslinker ratio is named Hydro-\SI{3}{\percent}.

\subsection{Indentation testing}
We measured the mechanical bulk modulus of hydrogels using an indentation testing method as performed and validated by others \cite{schulze2017polymer,chan2012spherical,oyen2014mechanical,lee2015three}. By indenting a soft sample with a spherical indenter and measuring its force-displacement response, we can apply Hertzian contact mechanics \cite{Johnson1985} to determine an elastic modulus. We used a custom-built indentation tester to perform these measurements (Fig.~S2). Samples were prepared in a cylindrical shape and oriented such that a flat surface was indented. All tests are completed within 15 minutes after removing the samples from the humidity-controlled chamber to ensure minimal weight loss from de-swelling to the ambient environment. Displacement speeds ranged from \SIrange{6}{10}{\milli\meter\per\minute}; slower or faster speeds did not affect the force-displacement curves, indicating the sample behaved quasi-statically and quasi-elastically, away from dynamic drainage and viscoelastic effects. The loading force, $F$, is proportional to the displacement of the ball bearing, $d$, raised to the power of $3/2$:

\begin{equation}
    F = \frac{4}{3}E^{*}R^\frac{1}{2}d^\frac{3}{2} \label{eqs1}\\
\end{equation}
where $R$ is the radius of the ball bearing in the setup and $E^{*}$ is an effective modulus. $E^{*}$ is related to the Young's moduli, $E$, and Poisson's ratios, $\nu$, of the sample and indenter such that
\begin{equation}
    \frac{1}{E^{*}} = \frac{1-\nu_\text{indenter}^2}{E_\text{indenter}} + \frac{1-\nu_\text{sample}^2}{E_\text{sample}}. \label{eqs2}
\end{equation}

The Poisson's ratio for all hydrogels, $\nu_\text{sample}$ was assumed to be $1/3$ as measured previously by others for crosslinked hydrogels \cite{wyss2010capillary,Geissler1981,Andrei1998}. Therefore, the bulk modulus is equivalent to the Young's modulus: $K_\text{sample} = E_\text{sample}$. Each sample was tested five times and the the force-displacement data was fit to Eq.~\ref{eqs1}, allowing us to determine $E_\text{sample}$ and its associated uncertainty in fitting. We identified two primary sources of uncertainty. The first being the standard deviation from five separate measurements, $\sigma$. The second being the average uncertainty from fitting $e = {\sqrt{\sum{e_i}^2}}/N$ where $e_i$ is a fitting uncertainty from a particular measurement and $N$ is the number of measurements. Thus, the total uncertainty in elastic modulus is $\sqrt{{\sigma}^2 + { e}^2}$, which can be visualized by the error bars in Fig.~\ref{fig.1}.

\begin{figure}
    \centering
    \includegraphics[width=0.5\linewidth]{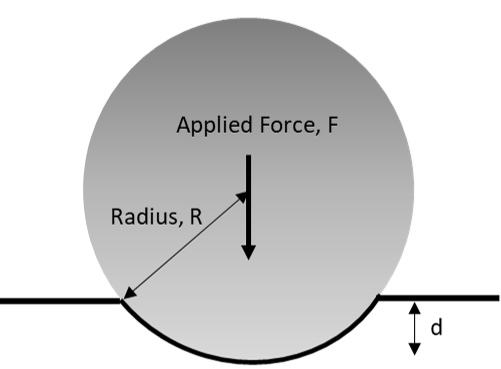}
    \caption{Scheme of indentation test based on contact machanics}
    \label{fig.5}
\end{figure}

\subsection{Humidity control}
To achieve stable relative humidities below \SI{100}{\percent}, we used a microfluidic controller (Elveflow) to mix dry and humid air flows (Fig.~S3). The dry air source was supplied by laboratory air supply at \SI{7}{\percent} RH while the humid air source was bubbled through DI water and achieved a humidity of around \SI{99}{\percent}. A custom-built PID control software was implemented to achieve humidity values of \SI{10}{\percent}, \SI{30}{\percent},\SI{50}{\percent},\SI{70}{\percent}, and \SI{90}{\percent}.

\subsection{FTIR Results}

Fourier-transform infrared (FTIR) spectroscopy (Shimadzu, IRSpirit, QATR-S) was used to verify that three different hydrogel polymer families (Pure, DMA, and Hydro) were synthesized. Within each family, FTIR spectra did not significantly change with crosslinker ratio. FTIR results are shown in Fig.~S4.

\section*{Conflicts of interest}
There are no conflicts to declare.

\section*{Acknowledgements}
It is a pleasure to acknowledge Mario R. Mata Arenales and Brandon Ortiz for helpful discussions, indentation testing, and humidity control. We also thank Suraj V. Pochampally and Jaeyun Moon for assistance with FTIR spectroscopy. This work was supported by the University of Nevada, Las Vegas through start-up funds, the Faculty Opportunity Award, the Top Tier Doctoral Graduate Research Assistantship program, and the Spring Semester Research Experience. SSD acknowledges support from the Princeton E-ffiliates Partnership of the Andlinger Center for Energy and the Environment, the Eric and Wendy Schmidt Transformative Technology Fund, Project X, and the Princeton Center for Complex Materials, a Materials Research Science and Engineering Center supported by NSF grant DMR-2011750.

\bibliographystyle{unsrtnat}
\bibliography{rsc}  






\end{document}


\title{Supplementary Information for "Scaling laws to predict humidity-induced swelling and stiffness in hydrogels"}

\maketitle
\renewcommand{\thefigure}{S\arabic{figure}}
\section{Chemicals}
The chemicals used in the preparation of hydrogels are listed below:

Acrylamide (AAm)

N,N' - Methylenebis(acrylamide) (MBA)

Ammonium Persulfate (APS)

N,N,N',N' - teramethylethane - 1,2 - dimine (TEMED)

N,N - Dimethylacrylamide (DMA)

All chemicals used in this paper are purchasd from Sigma-Aldrich Co.

\section{Figures}

\begin{figure*}
    \centering
    \includegraphics[width=\linewidth,keepaspectratio]{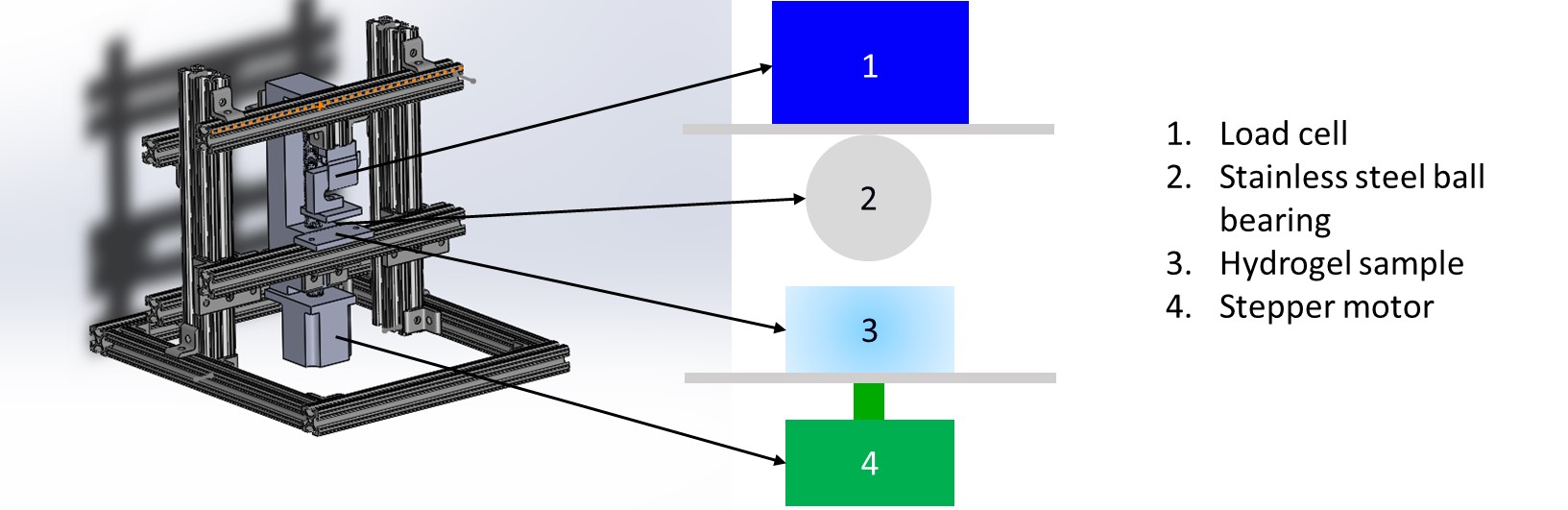}
    \caption{Indentation testing setup}
\end{figure*}

\begin{figure*}
    \centering
    \includegraphics[width=\linewidth]{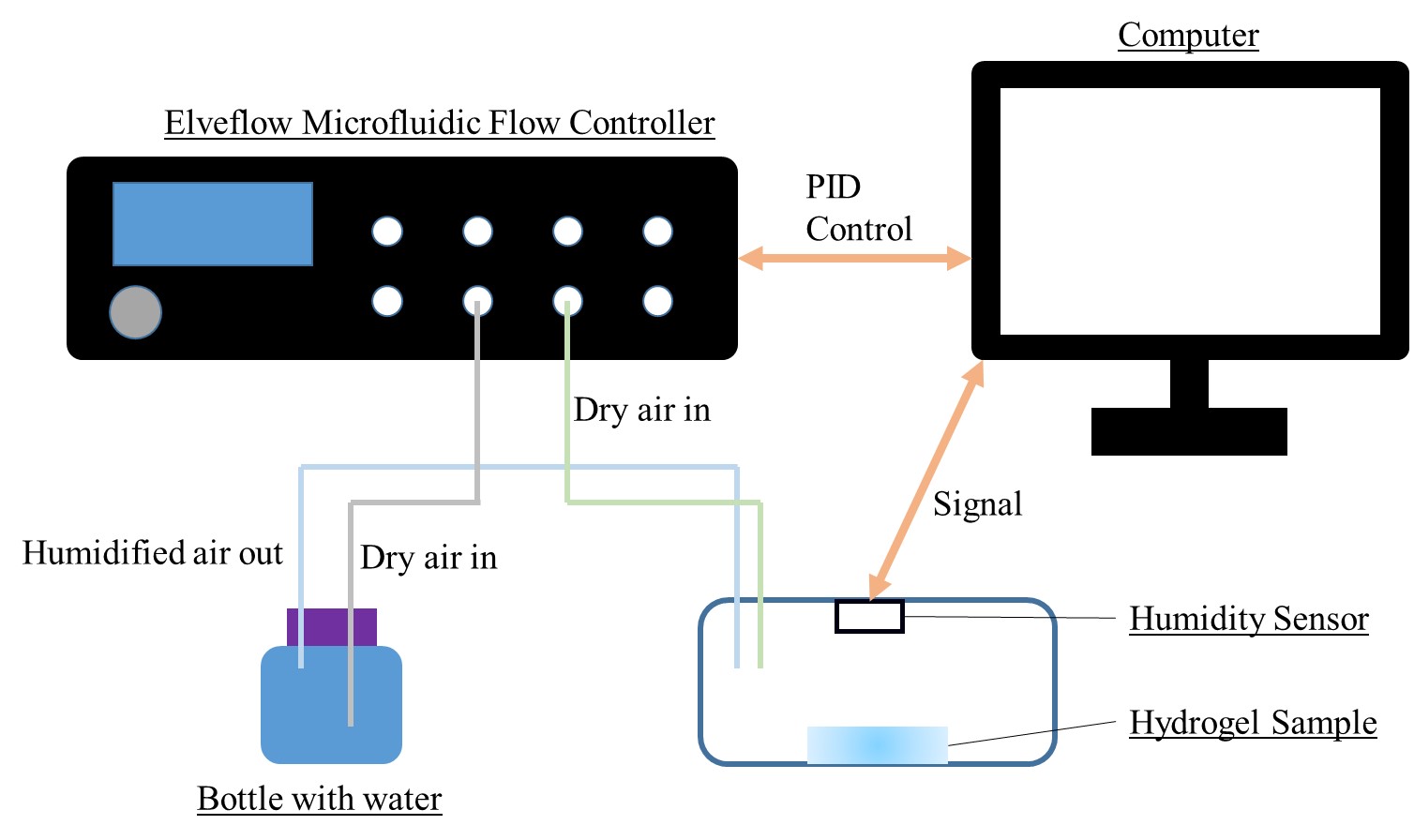}
    \caption{Humidity-controlled chamber setup}
\end{figure*}

\begin{figure*}
    \centering
    \includegraphics[width=\linewidth]{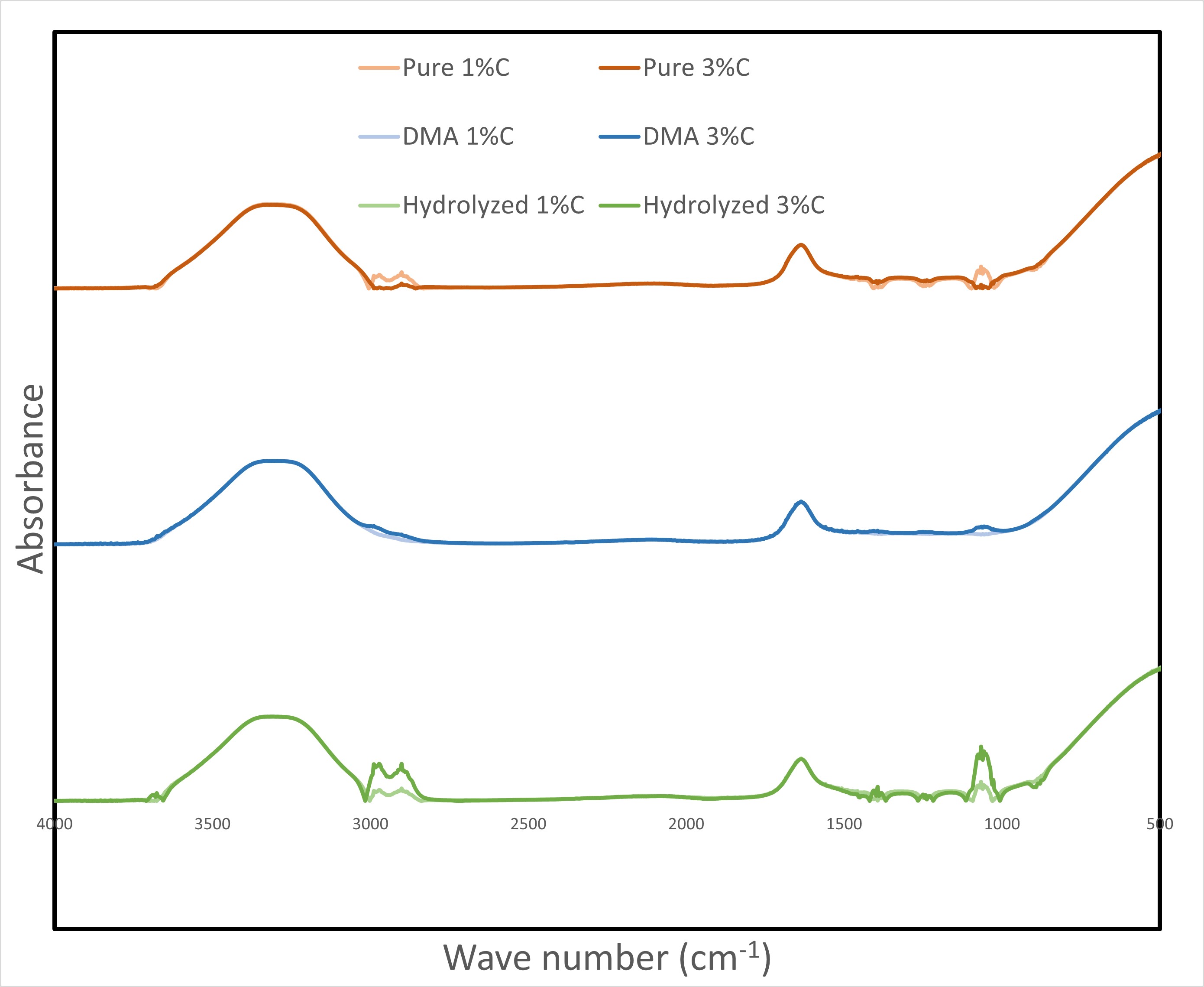}
    \caption{FTIR spectra indicate three different hydrogel polymer mesh families.}
\end{figure*}

\begin{figure*}
    \centering
    \includegraphics[width=\linewidth]{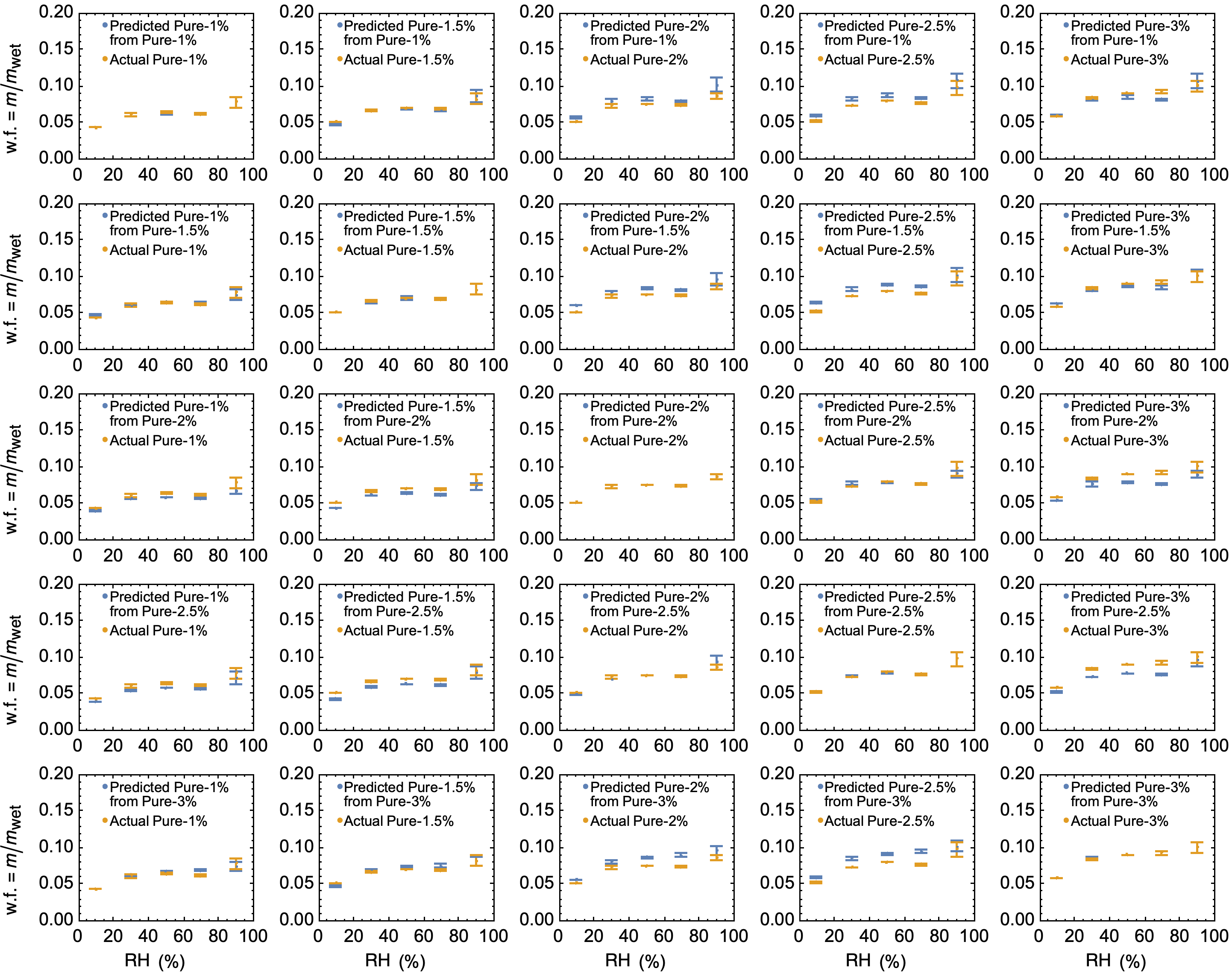}
    \caption{All prediction data (Pure PAAm Hydrogel) compared with actual data shows close agreement.}
\end{figure*}

\begin{figure*}
    \centering
    \includegraphics[width=\linewidth]{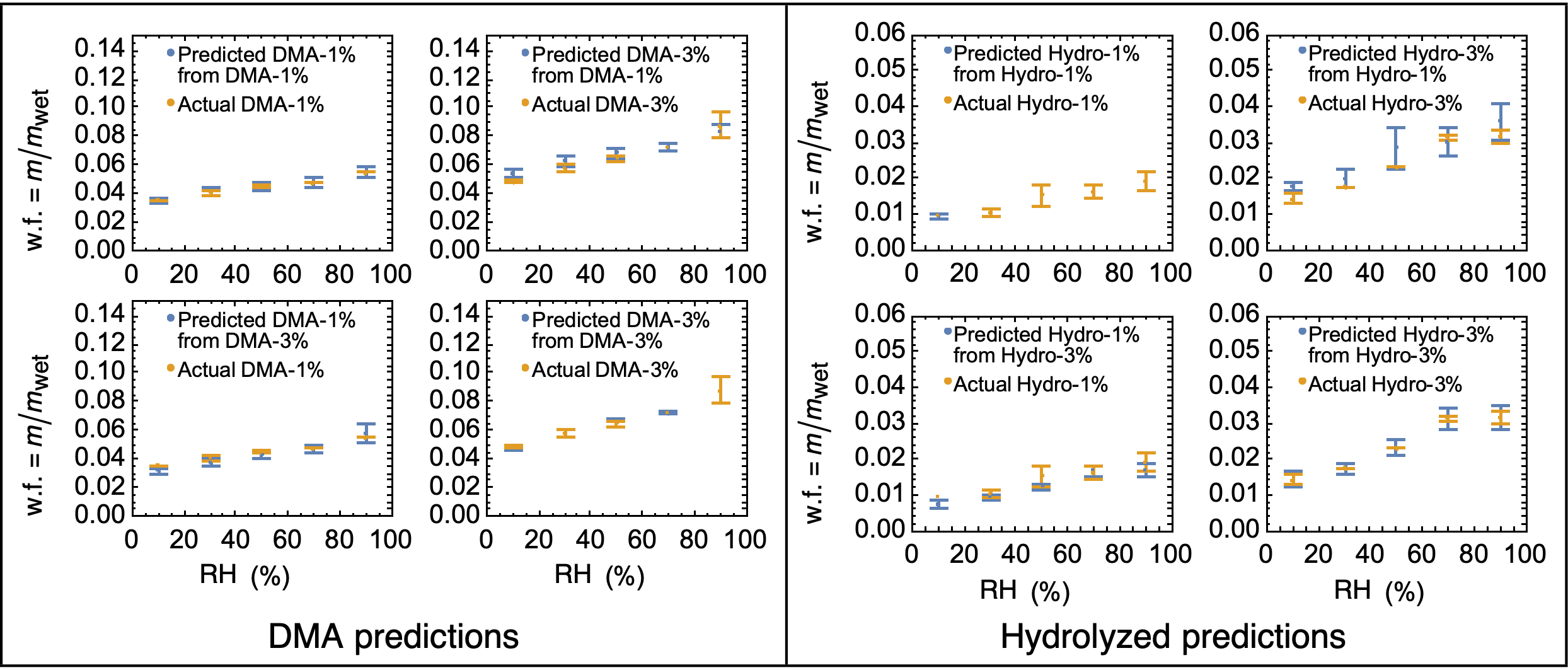}
    \caption{All prediction data (PAAm Hydrogel with DMA (left) or hydrolysis (right)) compared with actual data shows close agreement.}
\end{figure*}